# The incommensurate magnet iron monophosphide FeP: Crystal growth and characterization

I. O. Chernyavskii[a,b], S. E. Nikitin[c,d], Y. A. Onykiienko[c], D. S. Inosov[c,e], Q. Stahl[c], J. Geck[c,e], X. C. Hong[a], C. Hess[a], S. Gass[a], A. U. B. Wolter[a], D. Wolf[a], A. Lubk[a], D. V. Efremov[a], F. Yokaichiya[f], S. Aswartham[a,*], B. Büchner[a,c,e], I. V. Morozov[a,b,*]

[a] *Leibniz Institute for Solid State and Materials Research Dresden, Helmholtzstraße 20, D-01069 Dresden, Germany*
[b] *Department of Chemistry, Lomonosov Moscow State University, 119991 Moscow, Russia*
[c] *Institut für Festkörper- und Materialphysik, Technische Universität Dresden, D-01069 Dresden, Germany*
[d] *Max Planck Institute for Chemical Physics of Solids, Nöthnitzer Str. 40, D-01187 Dresden, Germany*
[e] *Würzburg-Dresden Cluster of Excellence ct.qmat, Technische Universität Dresden, 01062 Dresden, Germany*
[f] *Helmholtz Zentrum Berlin für Materialien und Energie, 14109 Berlin, Germany*

* corresponding authors:   morozov@inorg.chem.msu.ru, s.aswartham@ifw-dresden.de

## Abstract

We report an optimized chemical vapor transport method, which allows growing FeP single crystals up to 500 mg in mass and 80 mm$^3$ in volume. The high quality of the crystals obtained by this method was confirmed by means of EDX, high-resolution TEM, low-temperature single crystal XRD and neutron diffraction experiments. We investigated the transport and magnetic properties of the single crystals and calculated the electronic band structure of FeP. We show both theoretically and experimentally, that the ground state of FeP is metallic.  The examination of the magnetic data reveals antiferromagnetic order below $T_N$ =119 K while transport remains metallic in both the paramagnetic and the antiferromagnetic phase. The analysis of the neutron diffraction data shows an incommensurate magnetic structure with the propagation vector **Q** = (0, 0, ±δ), where δ ≈ 0.2. For the full understanding of the magnetic state, further experiments are needed. The successful growth of large high-quality single crystals opens the opportunity for further investigations of itinerant magnets with incommensurate spin structures using a wide range of experimental tools.

# 1. Introduction

Iron monophosphide FeP belongs to a large family of transition metal monopnictides, which crystallize in the MnP structure type (orthorhombic, space group *Pnma*). Besides FeP, this family includes CrP, MnP, CoP, WP, RuP, FeAs, CrAs, $Cr_{1-x}Mn_xAs$, and many other compounds which currently attract a lot of attention because of their nontrivial magnetic and superconducting properties [1-5]. The most interesting characteristic is the formation of a helical magnetic structure (HMS). So far, this property is found in four individual pnictides: CrAs, FeAs, MnP, and FeP [1]. Under pressure, the magnetic state can change or disappear. For instance, it was recently shown that in CrAs and MnP under high pressure, the helimagnetic behavior disappears and superconductivity occurs [3-5]. This has motivated physicists to a detailed study of the magnetic behavior of the sister compound FeP, in which HMS emerges below $T_N$ =120 K [6]. Despite considerable efforts, however, the details of the magnetic structure and mechanism of the HMS formation remain unclear [6–13].

The first HMS model for FeP was proposed in [6]. To explain the neutron diffraction data, the authors assumed a HMS with two different magnetic moments of the Fe atoms in the magnetic state despite a unique crystallographic position of Fe. To check this scenario, recently a series of studies on polycrystalline FeP were carried out by means of Mössbauer $^{57}$Fe spectroscopy, $^{31}$P nuclear magnetic resonance (NMR) spectroscopy, and X-ray photoemission spectroscopy [11-13]. The authors concluded that the complex Mössbauer spectrum can be explained by a model with a unique Fe magnetic moment in the intermediate spin state, provided the helical magnetic structure is strongly anisotropic. Additional information, which may resolve the real magnetic state, can be obtained from inelastic neutron scattering and $^{31}$P nuclear magnetic resonance, which requires sufficiently large crystals.

In this paper, we report the FeP crystal growth using the chemical vapor transport with iodine (ICVT) method. In general, this method is well established and is frequently used for the synthesis of single crystals of *d*-element phosphides [14]. In particular, this method has been used to obtain FeP single crystals. But the typical dimensions of FeP single crystals did not exceed 1−3 mm in all previous studies [6-10]. Here we give details of the optimized temperature regime, which allows growing large high-quality single crystals of sizes up to 1 cm and masses up to 0.3−0.5 g. As-grown single crystals were characterized by means of powder and single crystal x-ray diffraction and SEM equipped with EDX analyzer. For the largest crystals, their single crystallinity was confirmed by the Laue method. Neutron diffraction measurements on one of the largest single crystals show very sharp resolution-limited Bragg reflections evidencing the single domain character, which further confirms the quality of the as-grown crystals.

Further confirmation of the high quality stems from high-resolution TEM measurements on a lamella prepared from the as-grown material, and from the high residual resistivity ratio (RRR) value over 560. Magnetic susceptibility and resistivity results on our single crystals are in good agreement with the literature data [9, 10], evidencing an antiferromagnetic transition at about 120 K.

## 2. Crystal growth

All procedures for the preparation of the reaction mixture were performed in a glove box in an atmosphere of dry argon. To carry out experiments on the single crystal growth, a sample of stoichiometric polycrystalline iron phosphide (FeP) was first prepared in accordance with [12]. Iron powder (Alfa Aesar, 99.995%) and pieces of red phosphorus (Alfa Aesar, 99.999%) were mixed stoichiometrically by grinding in a mortar. The resulting mixture was placed in a niobium foil container and sealed in an evacuated quartz ampoule. The ampoule was slowly heated (30 °C/h) to the temperature of 850 °C and annealed at this temperature for 48 h. Then, a series of experiments on the growth of FeP single crystals with ICVT were carried out. Up to 0.5−2 g of polycrystalline FeP and 0.03−0.2 g of $I_2$ (for the creation of a concentration of 0.2−18 mg/ml iodine) were placed in a quartz ampoule with a length of 10−20 cm and an inner diameter of 1.2−3 cm. The ampoule was then evacuated to a pressure of $10^{-5}$ mbar and sealed. To avoid significant losses of iodine, after five minutes of pumping at room temperature, the ampoule was cooled down with liquid nitrogen from the bottom side. Then the sealed ampoule was placed in a two-zone furnace. In different experiments the charge zone temperature was changed in the range of 780–850 °C, the temperature difference between the charge and the growth zone was changed in the range from 30 to 150 °C. The duration of the experiments was in between 7 and 30 days.

The best results were obtained using the following conditions: ampule length 12 cm, inner diameter 1.2 cm, mass of polycrystalline FeP and $I_2$ in the charge zone — 1 g and 0.1 g, respectively. The total heating time was 30 days. The temperature of the charge zone ($T_1$) was 850 °C throughout the synthesis. The temperature of the growth zone was changed in accordance with the following mode: annealing at 950 °C for 10 hours, then cooling to 850 °C for 5 hours. After that the temperature in the growth zone was gradually reduced during one month to 780 °C and the oven was turned off. The ampoule was cut, and the resulting single crystals were removed. On some crystals the surface was contaminated with condensed gaseous byproducts during the transport (according to Ref., [15], these can be: $I_2$, $PI_3$, $P_4$, $FeI_2$). These crystals were subsequently washed with distilled water and acetone to remove these condensed phases.

With these optimized conditions we were able to reproduce single crystals of FeP with a mass of up to several tenths of a gram. As a rule, in the growth zone there were 6−10 single crystals, among which the mass of the largest was 0.3−0.5 g, and the remaining single crystals had a mass of

0.05−0.1 g and less. A typical view of the single crystals obtained is shown in Fig. 1. Bulk single crystals contain many as grown facets and have a metallic luster.

## 3. Experimental methods

The actual composition of the crystals was determined by using an x-ray energy dispersive spectrometer (INCA X-sight, Oxford Instruments) mounted on a field emission scanning electron microscope JEOL JSM 6490 LV with a W-cathode. A quantitative analysis of the spectra was performed using the INCA software (Oxford Instruments). According to the data obtained, the composition of single crystals is $Fe_{0.52(2)}P_{0.48(2)}$ (Fig. 2). Taking into account the accuracy of the EDX, this result corresponds to the stoichiometric composition FeP.

X-ray powder diffraction data were collected using a Huber diffractometer with $CoK_\alpha$-radiation in the transmission mode. An XRD pattern of a polycrystalline sample obtained by grinding a single crystal confirmed that the substance obtained is iron monophosphide. The refined unit cell parameters (sp. gr. *Pnma*, a = 5.1738(4), b = 3.0877(4), c = 5.7718(6), Å V = 92.21(2) Å$^3$) are in good agreement with the literature data [16]. For the largest single crystals, their single crystallinity was confirmed by the Laue method (Fig. 3).

The single-crystal x-ray diffraction data in the temperature range 300 K to 30 K were collected on a Bruker-AXS KAPPA APEX II diffractometer with graphite-monochromated Mo $K_\alpha$-x-ray radiation, equipped with an Oxford Cryosystems N-HeliX low temperature device. The distance between sample and CCD-detector was set to 45.1 mm. For the diffraction experiments the detector was positioned at 2θ positions of 30° and 45° for the measurements between 100 and 300 K and 30° for the data acquisition at 30 K, using an ω-scan mode strategy at four different φ positions (0°, 90°, 180° and 270°). For each run 240 frames with Δω = 0.5° were measured. All data processing was performed in the Bruker APEX3 software suite [17], the reflection intensities were integrated using SAINT [18] and a multi-scan absorption correction was applied using SADABS [19]. The subsequent weighted fullmatrix least-squares refinements on F$^2$ were done with SHELX-2012 [20] as implemented in the WinGx 2014.1 program suite [21].

High-resolution TEM investigations were performed on a thin electron transparent lamella oriented in [001]-direction, which was prepared from the as-grown single crystals with the Focused Ion Beam (FIB) technique. HR-TEM images were acquired by using a double-corrected Titan$^3$ 80-300 instrument (ThermoFisher Company, USA), operated at an acceleration voltage of 300 keV. After correcting for geometric aberrations, a spatial resolution of approximately 0.08 nm is obtained.

Temperature and field dependent magnetization studies were performed using a commercial Superconducting Quantum Interference Device (SQUID) magnetometer by Quantum Design

(SQUID-VSM). A background subtraction to remove parasitic signals from the glue and the sample holder was applied. The measurements were performed for zero-field cooled (ZFC) and field-cooled (FC) conditions and for two crystallographic directions, i.e., parallel and perpendicular to the growth direction of the crystal, which approximately corresponds to the crystallographic *b* direction.

The electrical resistivity measurements were carried out by using a standard four-point technique in a home-built dip stick from room temperature down to 8 K. Silver wires were electrically contacted to the sample surface with silver paint. The current was set to 2 mA and the direction alternated with low frequency. The sample dimensions were measured with a Zeiss Stemi 2000-C stereo microscope. The crystal was 1.13 mm in length (0.61 mm distance between potential contacts), 0.37 mm in width and 0.23 mm in thickness.

## 4. Results and discussion

### 4.1. Crystal growth

A comparison of the conditions for FeP crystal growth by ICVT, which were used in this work, with the ones given in the literature, shows that most of the parameters (ampoule size, iodine concentration, temperature in the charge and growth zone) do not differ much [7-10, 14, 15]. For example, our experiments confirm the conclusions of [14] about the need to use a sufficiently large amount of iodine. Indeed, in experiments where the iodine concentration was about 2−3 mg/ml and less, the transport rate was insignificant. In addition, in several experiments, we observed not only FeP single crystals, but also some amounts of iron iodide $FeI_2$ in the growth zone. This observation is consistent with the conclusions of the authors of [15], who, based on theoretical calculations, indicated the possibility of co-crystallization of FeP and $FeI_2$ under certain conditions. The crystallization of metal iodide $MI_2$ together with phosphide *M*P was also observed for *M* = Mn, Cr and Co [14] in experiments carried out at lower mean temperatures (in the case of *M* = Co with T ≲ 800°C [22]) and with sufficiently high amounts of iodine.

The key finding, which made it possible to grow large single crystals, is the applied temperature regime. We have found that a slow decrease of the temperature in the growth zone leads to the growth of FeP single crystals of the size of 1 cm. The reason is that the size of the single crystals is determined by the ratio of the nucleation rate and the crystal growth rate [23]. At low degrees of supersaturation (i.e., when temperatures in the growth zone and change zone are close, the pressure of the reactants in the growth zone only slightly exceeds the equilibrium values), the rate of nucleation is small compared with the growth rate of a single crystal. As a result, only a few viable nucleation centers are formed that begin to grow. If a further decrease of the temperature in the growth zone occurs sufficiently slowly, then the growth rate of a single crystal may remain higher than the rate of nucleation due to an increase in the surface area of the single crystal on which crystallization takes place. With a decrease in temperature, a crystallization of iron iodide $FeI_2$ begins to play a certain role, as a result, a part of iodine is removed from the system, thereby reducing the degree of supersaturation, and the possibility of additional nucleation.

### 4.2. High-resolution TEM investigations

The atomic resolution images of the TEM lamella oriented in [001]-orientation have been recorded under parallel illumination (Fig. 4a), from which the diffractograms (i.e. Fourier transforms of the image intensity) have been computed in a second step (Fig. 4b). The distances and symmetry of systematic reflections observed in the diffractograms correspond well to the crystal structure determined in [16] (Fig. 4c). The same holds for the [001]-projection of the crystal structure observed in the high-resolution image. Note, however, that the origin of the unit cell with respect to the high-resolution image cannot be determined unambiguously due to dynamical scattering and residual aberrations.

### 4.3. Magnetic and transport properties

Magnetic susceptibility measurements were performed on a single crystal of FeP in an external magnetic field of 1 T parallel and perpendicular to the main growth direction of the crystal, which roughly correspond to parallel and perpendicular to the b direction, respectively (Fig. 5 a). Our results are in good agreement with the literature data [9, 10], evidencing an antiferromagnetic transition at $T_N = 120$ K. No splitting between ZFC and FC curves has been observed. Interestingly, the susceptibility data still show some anisotropy in the paramagnetic region above $T_N$, together with a very broad maximum around 220 K, which has been claimed to arise from an anisotropic, low-dimensional chain structure of FeP giving rise to short-range correlations above the Néel temperature along the Fe zigzag chain [9]. Field-dependent magnetization studies down to 2 K show a perfectly linear field dependence for both orientations and an absence of any metamagnetic transition up to 7 T (not shown). Note that in comparison to other works on single and polycrystalline FeP [8, 10], the rather constant susceptibility at low temperature shown in Fig. 5a indicates the absence of paramagnetic impurities and highlights the high quality of the large single crystals used in our study.

The absolute value of the resistivity and the shape of the curve shows good agreement with the literature data [9]. The temperature dependence of the resistivity shows a kink at around 120 K, which corresponds well to $T_N$ (Fig. 5b). The relatively steeper decrease of the resistivity below $T_N$ is characteristic for reduced scattering in a spin-density-wave ordered phase, which is possible only in crystals with high purity, and resembles similar findings in FeAs and iron-based superconductors [24, 25]. The purity of the crystal is further confirmed by the large RRR of over 560.

### 4.4. The neutron scattering investigation

Single-crystal neutron diffraction data presented in Fig. 6 were measured using the E2 diffractometer at HZB, Berlin. The sample was mounted in the (*h* 0 *l*) scattering plane on an Al sample holder and placed in a cryostat. The neutron wavelength from a pyrolytic graphite (002) monochromator was fixed at λ = 2.41 Å ($k_i$ = 2.61 Å$^{-1}$), and a 30′ radial collimator was installed after the sample. Low-temperature neutron diffraction data in the left panel of Fig. 6 show the appearance of incommensurate magnetic satellites around some of the structural Bragg reflections, such as (0 0 2±δ) and (1 0 1±δ), with δ ≈ 0.2. These satellites disappear above the magnetic ordering temperature, $T_N$. The magnetic and structural reflections appear equally sharp and resolution-limited, which is indicative of a long-range magnetic order. The magnetic ordering wave vector is consistent with a spin spiral propagating along the (001) direction, reported in earlier works [6].

### 4.5. Temperature-dependent singe-crystal x-ray diffraction investigation

To obtain detailed information on the average structure, single crystal x-ray diffraction measurements were performed on a piece of about 50 μm in diameter isolated from an as-grown crystal. The low merging *R*-factors (about 3%) and the sharp reflection pattern indicate the high quality of the crystals (Fig. 7). Table 1 summarizes parameters of the data collection and the results of the structural refinement. The atomic positions, isotropic and anisotropic displacement parameters are listed in Tables 2 and 3.

In agreement with previously reported diffraction experiments, iron monophosphide crystallizes in the orthorhombic space group *Pnma* or MnP-type structure [16]. However, also the non-centrosymmetric *Pn*2$_1$*a* symmetry was suggested as the appropriate space group, both space groups are indistinguishable from the analysis of the systematic extinctions alone [7]. In order to address this question, we performed crystallographic refinements in both space groups for all datasets within the whole temperature range investigated by single crystal x-ray diffraction (30 – 300 K). In accordance with most of the former studies, the centrosymmetric *Pnma* symmetry provides the most accurate model, with the Fe and P atoms lying in a crystallographic mirror plane perpendicular to the *b* axis at *y* = ±0.25. More specifically, refinements with anisotropic atomic displacement parameters converged in *Pnma* to final w$R_2$ values below 5% and in *Pn2$_1$a* to final w$R_2$ values above 5% (always referring to all data). Also, the residual electron density maxima in the centrosymmetric model are about 0.2 e·Å$^{-3}$ lower than in the non-centrosymmetric *Pn2$_1$a* space group.

Interestingly, the high quality of our single crystals enabled observation of very weak additional reflections of the type *hk*0 with *h* = odd, which are highlighted by red rings in Fig. 7. These peaks are by a factor of 100 to 1000 less intense than the Bragg peaks with *h* = even in the *hk*0 layer. Nevertheless, the *hk*0 reflections with *h* = odd can clearly be detected at all studied temperatures. *hk*0 reflections with *h* = odd, which have not been included in the refinements described above, violate the extinction rule for the MnP-type structure and encourage further research.

**4.6. Ab initio calculation of the electronic band structure in the paramagnetic state**

To shed light on the electronic band structure close to the Fermi level in the nonmagnetic state, we have used density functional theory (DFT). In the calculations, the crystal structure reported in [16] was assumed. The full relativistic generalized gradient approximation (GGA) in the Perdew-Burke-Ernzerhof variant was used for the exchange correlation potential implemented in the Full Potential Local Orbital band structure Package (FPLO) [26, 27]. A **k** mesh of 12×12×12 **k**-points in the whole Brillouin zone was employed. In Fig. 8, the total and partial density of states (DOS, PDOS) projected on each atomic species are presented. DFT results suggest a metallic state for FeP as indicated by a finite density of states (DOS) at the Fermi level $E_F$, what is in full agreement with our resistivity results. From −2.5 to 2 eV, the Fe 3*d* states represent the main contribution to the DOS, together with a contribution from P *p* states. For states ranging from −8 to −2.5 eV and from 2 to 4.5 eV in energy, both Fe 3*d* and P 3*p* states are contributing to the DOS. Fig. 9 shows the calculated band structure along some high-symmetry directions computed with the method outlined above. In this range of energy −2 to 2 eV, the bands are made of Fe states together with a contribution from P states, as seen on the PDOS plots.

**5. Conclusions**

We have demonstrated that growth of single crystals by means of chemical vapor transport with iodine as a transport agent can be optimized by choosing a corresponding temperature regime. As a result, high-quality single crystals of FeP up to 500 mg in mass and 80 mm$^3$ in volume were grown.

The quality of the crystals was confirmed by means of EDX, XRD, high-resolution TEM, magnetic and transport studies.

For the largest crystals, their single crystallinity and high quality was confirmed by the Laue method, as well as by single crystal neutron diffraction. The analysis of the low-temperature neutron-diffraction data yields an incommensurate magnetic structure with the propagation vector **Q** = (0, 0, ±δ), where δ ≈ 0.2, emerging below the Néel temperature $T_N$ =120 K. The experimental investigation is supplemented by a theoretical calculation of the electronic band structure in the paramagnetic state, which is in agreement with the metallic state of the compound. To explain the formation of a double helicoid, a low-temperature single crystal XRD study of FeP was performed for the first time. This investigation confirmed the absence of a phase transition, which would lead to the formation of two crystallographically independent positions of the Fe atom. However, weak XRD-intensity has been observed at forbidden *hk0*-positions, indicative of a broken glide plane symmetry. This may indicate different independent Fe-positions. Based on the available data, however, we cannot pin down the precise cause of the broken symmetry. This will require detailed further study, which is beyond the scope of this manuscript.

Another possible reason for the formation of a double helicoid is the pronounced anisotropy of the magnetic moment of the Fe atoms. However, the calculation of the electronic structure did not reveal the presence of such anisotropy. Thus, further studies are needed to shed light on the cause of the unusual magnetic behavior of FeP. The successful growth of high-quality single crystals opens an opportunity for further investigations of itinerant helimagnets. Currently, the obtained single crystals are already being studied by $^{31}$P NMR and inelastic neutron-scattering methods. Preliminary results show that, thanks to the sufficiently large size and quality of the obtained single crystals, it is possible to obtain new intriguing information about the magnetic behavior of this compound and the entire family of helimagnets with the MnP structure.

## Acknowledgements

We are grateful for the helpful discussion Dr. Vladimir Aleshin. We thank Y. V. Tymoshenko and P. Y. Portnichenko for the assistance with sample alignment and neutron data processing. We thank U. Nitzsche for technical support. I.V.M, S.A., B.B. and D.V.E thank DFG and RSF for financial support in the frame of the joint DFG-RSF project "Weyl and Dirac semimetals and beyond −prediction, synthesis and characterization of new semimetals". IC acknowledges the RFBR grants No. 17-52-80036 and No. 18-33-01282. S.A. acknowledges financial support from the German Research Foundation (DFG) under the grant No. AS 523/4-1. D.S.I. acknowledges financial support from the German Research Foundation (DFG) under the grant No. IN 209/4-1, and D.S.I., B.B. and A.U.B.W. from the projects C03, C06, C09, B01 of the Collaborative Research Center SFB 1143 at the TU Dresden (project-id 247310070). S.E.N. acknowledges support from the International Max Planck Research School for Chemistry and Physics of Quantum Materials (IMPRS-CPQM). C.H. and X.H. acknowledge support by the European Research Council (ERC) under the European

Unions' Horizon 2020 research and innovation program (Grant Agreement No. 647276-MARS-ERC-2014-CoG). A.L. and D.W. acknowledge support by the European Research Council (ERC) under the Horizon 2020 research and innovation program of the European Union (Grant Agreement No. 715620). B.B. acknowledges support from the German Research Foundation (DFG) in the framework of the Würzburg-Dresden Cluster of Excellence on Complexity and Topology in Quantum Matter – *ct.qmat* (EXC 2147, project-id 390858490).

**Table 1.** Details on data collection and structure refinement of FeP as determined from single-crystal X-ray diffraction.

| | 300 K | 200 K | 100 K | 30 K |
|---|---|---|---|---|
| Crystal data | | | | |
| *Space group* | *Pnma* | *Pnma* | *Pnma* | *Pnma* |
| *a* (Å) | 5.197(3) | 5.1927(10) | 5.1950(12) | 5.195(4) |
| *b* (Å) | 3.0994(15) | 3.0870(6) | 3.0802(7) | 3.083(3) |
| *c* (Å) | 5.794(3) | 5.7884(11) | 5.7909(13) | 5.797(5) |
| *V* (Å$^3$) | 93.32(8) | 92.79(3) | 92.66(4) | 92.85(13) |
| $M_r$ | 86.82 | 86.82 | 86.82 | 86.82 |
| *Z* | | | | |
| $\rho_{calc}$ (g/cm$^3$) | 6.179 | 6.215 | 6.223 | 6.211 |
| $\mu$ (mm$^{-1}$) | 16.716 | 16.813 | 16.835 | 16.802 |
| Data collection | | | | |
| 2 Θ *max* (°) | 62.950 | 62.982 | 62.004 | 62.960 |
| *Absorption correction* | Multi-Scan | Multi-Scan | Multi-Scan | Multi-Scan |
| $T_{min.}$ | 0.2110 | 0.2098 | 0.2097 | 0.2243 |
| $T_{max.}$ | 0.3279 | 0.3279 | 0.3279 | 0.3259 |
| $N_{measured}$ | 2003 | 2012 | 2005 | 1017 |
| $N_{independent}$ | 180 | 180 | 179 | 179 |
| $R_{int}$ (%) | 3.14 | 3.00 | 2.71 | 2.52 |
| Refinement | | | | |
| $N_{parameters}$ | 14 | 14 | 14 | 14 |
| $R_1 > 4s$ (%) | 1.63 | 1.74 | 1.53 | 1.52 |
| $R_1$ all (%) | 1.71 | 1.82 | 1.57 | 1.64 |
| $wR_2 > 4s$ (%) | 4.25 | 4.72 | 4.12 | 3.88 |
| $wR_2$ all (%) | 4.31 | 4.78 | 4.16 | 3.94 |
| G.O.F. | 1.196 | 1.217 | 1.159 | 1.156 |
| $\Delta\rho_{min}$ (e/Å$^3$) | -0.798 | -0.802 | -0.782 | -0.779 |
| $\Delta\rho_{max}$ (e/Å$^3$) | 0.515 | 0.619 | 0.572 | 0.542 |
| *weights w* (a,b) | 0.0290 | 0.0327 | 0.0295 | 0.0238 |
| | 0.0000 | 0.0000 | 0.0000 | 0.0000 |
| *Extinction k* | 0.65 | 0.51 | 0.34 | 0.33 |

$$T_{min} = minimum Transmission, T_{max} = maximum Transmission, R_1 = \sum||F_0|-|F_c||/\sum|F_0|,$$

$$wR_2 = \{\sum[w(F_0^2 - F_c^2)^2]/\sum[w(F_0^2)^2]\}^{1/2}, \ w = 1/\left[\left(\sigma^2(F_0^2)\right) + (aP)^2 + bP\right]$$

where $P = [2F_c^2 + max(F_0^2, 0)]/3$

**Table 2.** Atomic coordinates and equivalent isotropic displacement parameters of FeP single crystals at selected temperatures.

|  | 300 K | 200 K | 100 K | 30 K |
|---|---|---|---|---|
| Fe |  |  |  |  |
| x | 0.00194(5) | 0.00197(5) | 0.00192(5) | 0.00188(5) |
| y | 0.25 | 0.25 | 0.25 | 0.25 |
| z | 0.20040(6) | 0.20047(6) | 0.20060(6) | 0.20068(6) |
| $U_{eq}$ | 0.00518(19) | 0.00390(20) | 0.00292(18) | 0.00257(16) |
| P |  |  |  |  |
| x | 0.19145(9) | 0.19132(10) | 0.19114(9) | 0.19131(10) |
| y | 0.25 | 0.25 | 0.25 | 0.25 |
| z | 0.56851(9) | 0.56815(10) | 0.56781(9) | 0.56776(10) |
| $U_{eq}$ | 0.00498(20) | 0.00393(22) | 0.00310(19) | 0.00302(19) |

**Table 3.** Anisotropic displacement parameters of FeP single crystals at selected temperatures.

|  | 300 K | 200 K | 100 K | 30 K |
|---|---|---|---|---|
| Fe |  |  |  |  |
| $U_{11}$ | 0.00428(25) | 0.00413(27) | 0.00332(23) | 0.00334(22) |
| $U_{22}$ | 0.00736(25) | 0.00488(27) | 0.00324(23) | 0.00312(23) |
| $U_{33}$ | 0.00389(27) | 0.00269(30) | 0.00220(26) | 0.00126(24) |
| $U_{13}$ | -0.00002(8) | -0.00009(8) | 0.00006(8) | -0.00006(9) |
| P |  |  |  |  |
| $U_{11}$ | 0.00542(27) | 0.00497(30) | 0.00388(26) | 0.00391(26) |
| $U_{22}$ | 0.00521(28) | 0.00354(31) | 0.00314(27) | 0.00317(28) |
| $U_{33}$ | 0.00430(32) | 0.00329(36) | 0.00229(31) | 0.00199(31) |
| $U_{13}$ | 0.00031(16) | 0.00001(17) | 0.00033(16) | -0.00011(18) |

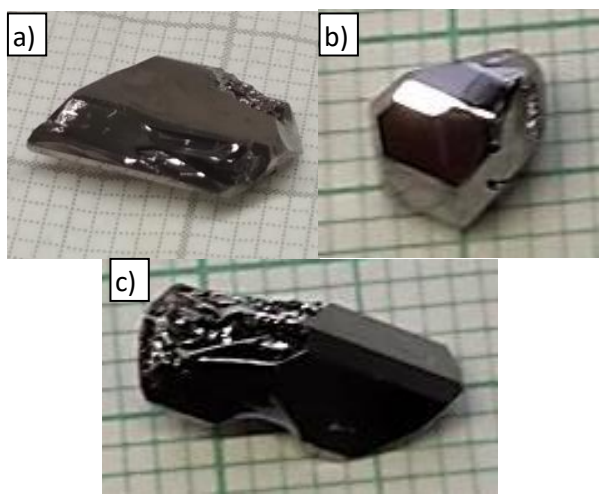

Fig. 1. Examples of FeP single crystals grown by the iodide vapor transport method. (a) - This crystal (m=519 mg) was used for inelastic neutron scattering experiments, (b, c) – the heaviest crystals from other experiments (407 and 346 mg, respectively). Cell scale is 1 mm.

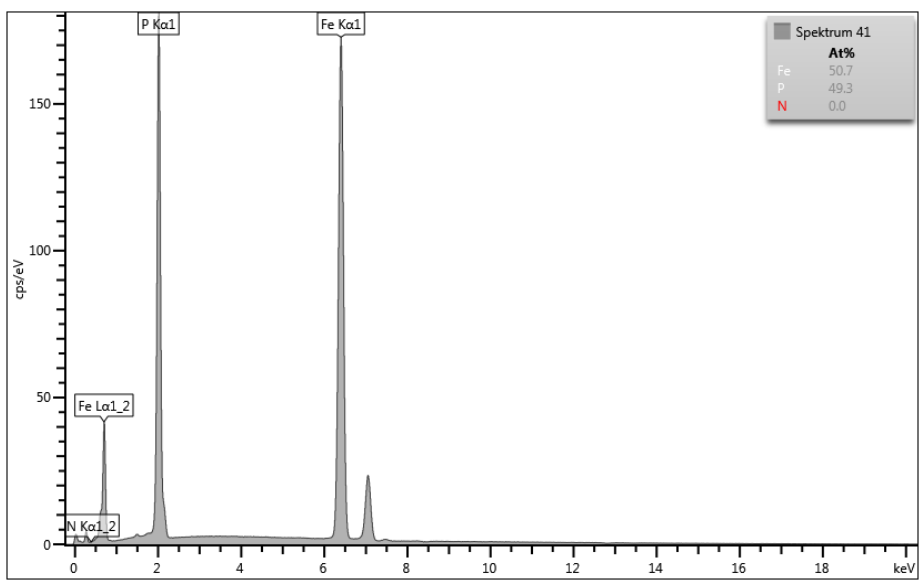

Fig. 2. EDX spectrum of a FeP single crystal.

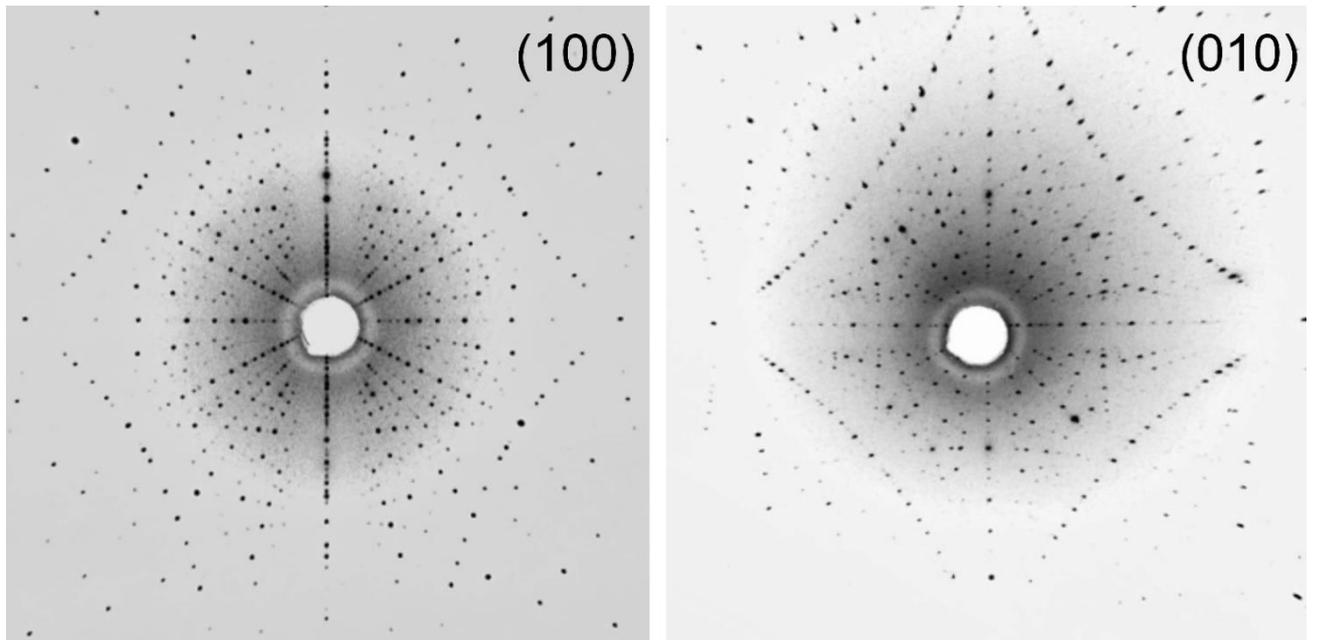

Fig. 3. X-ray Laue diffraction pattern of FeP crystal (see fig. 1b), measured along the *a* and *b* axes.

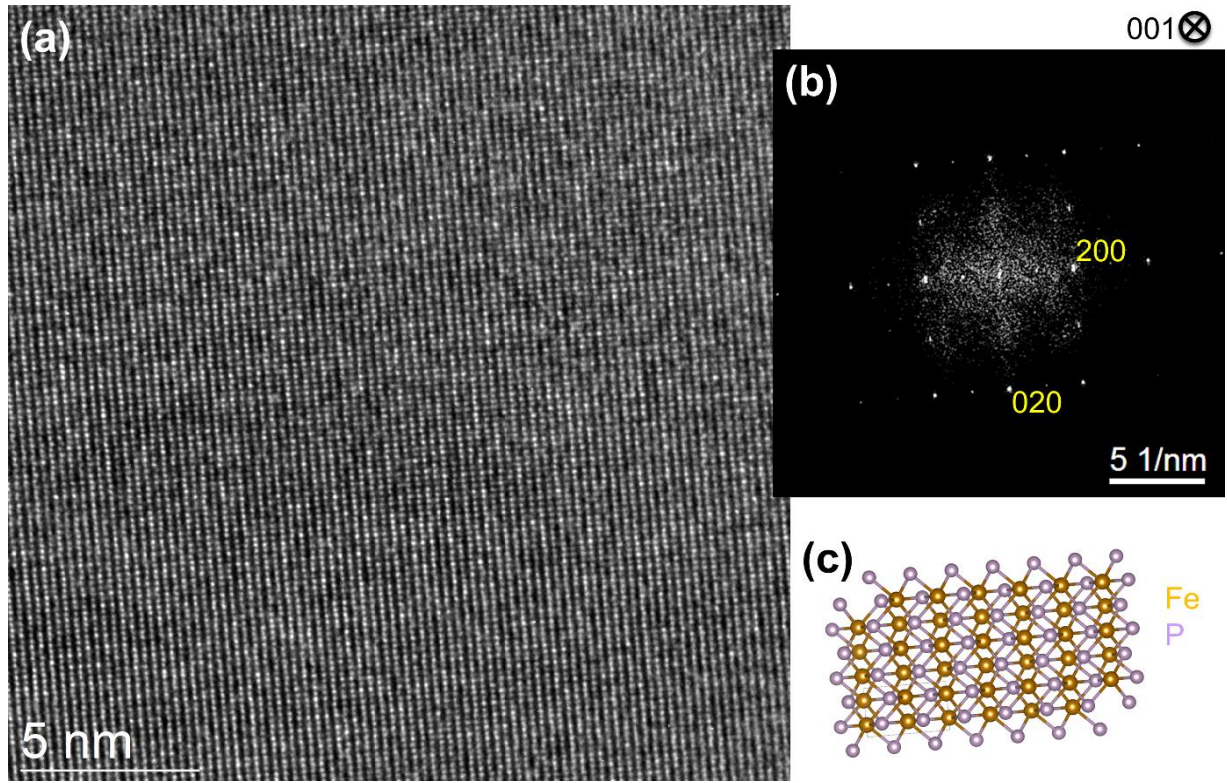

Fig. 4. High-resolution transmission electron microscopy (HRTEM) of FeP. (a) HRTEM image recorded in [001]-orientation. (b) Fourier transform (diffractogram) computed from the HRTEM image (a). (c) Atomic structure model (Fe gold, P violet color).

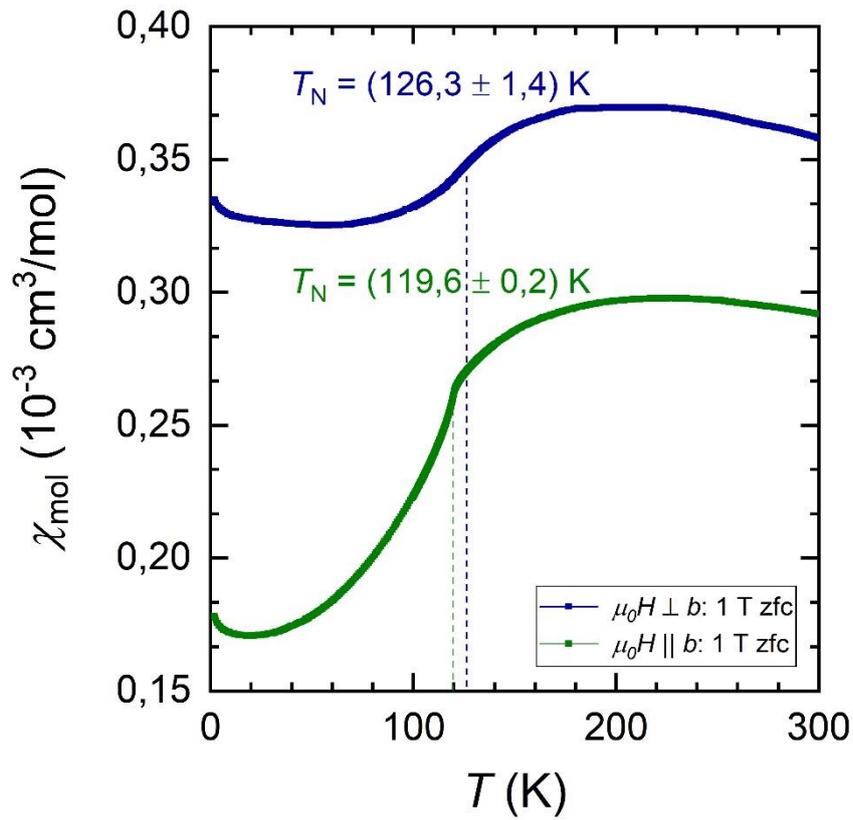

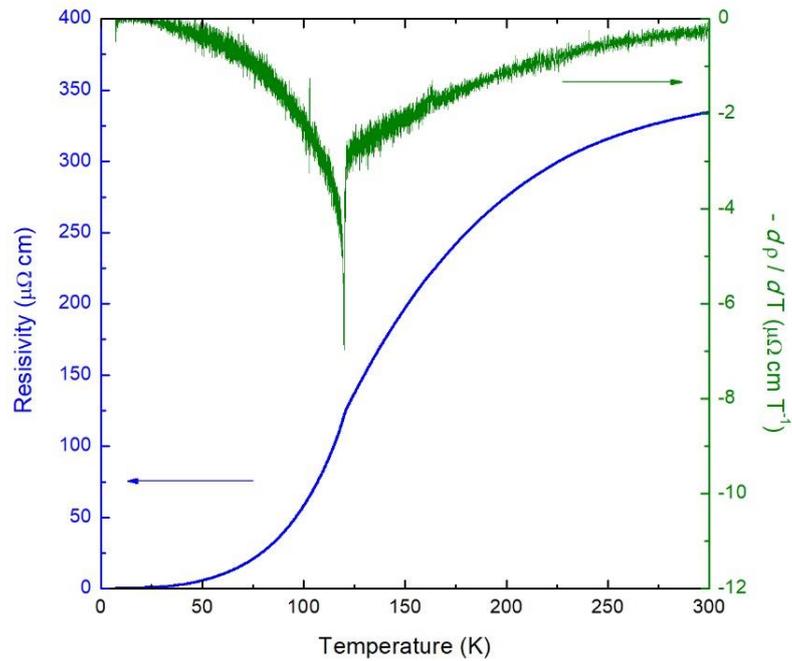

Fig. 5. Magnetic and transport properties of single crystal FeP. (a) the temperature dependence of the magnetic susceptibility of FeP; (b) The temperature dependence of the resistivity shows a kink at around 120 K, which is seen more clearly in its temperature derivative (right axis).

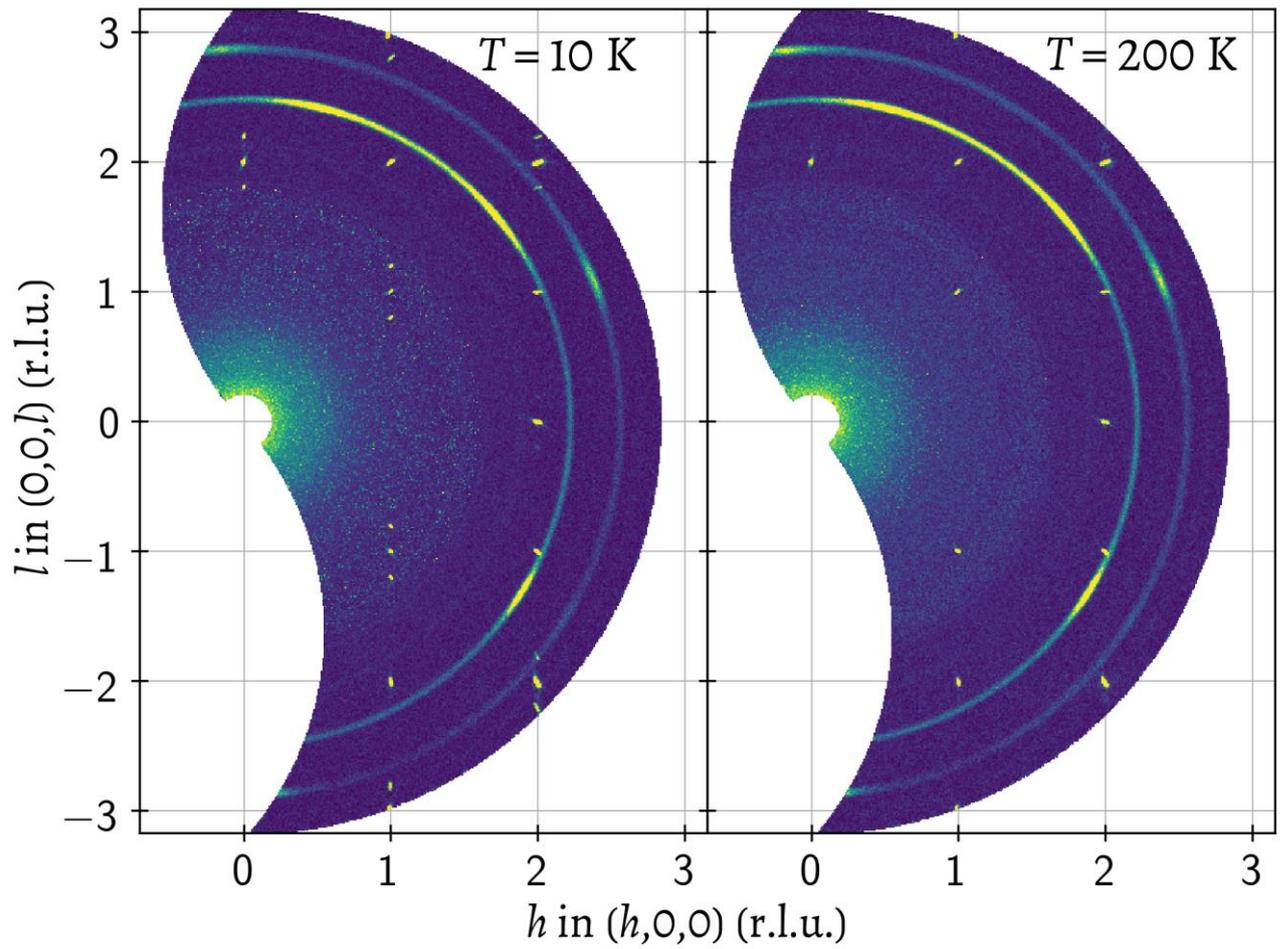

Fig. 6. Neutron diffraction intensity maps in the ($h$ 0 $l$) plane, measured at $T$ = 10 K (left) and $T$ = 200 K (right). In the low-temperature dataset, incommensurate magnetic satellites are seen around the (101), (002), (202) and equivalent Bragg reflections.

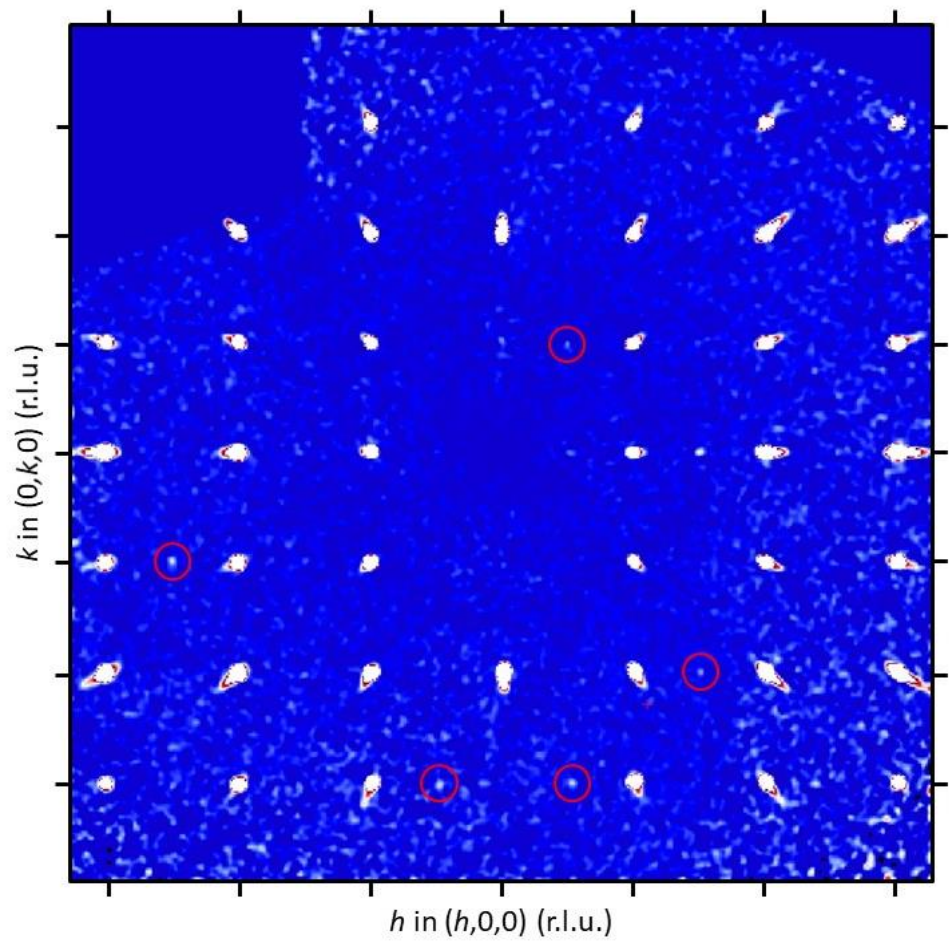

Fig. 7. X-ray diffraction intensity map in the (*hk*0) plane, measured at T = 300 K. The Bragg peaks violating the extinction rules for the space group *Pnma* are marked with red circles.

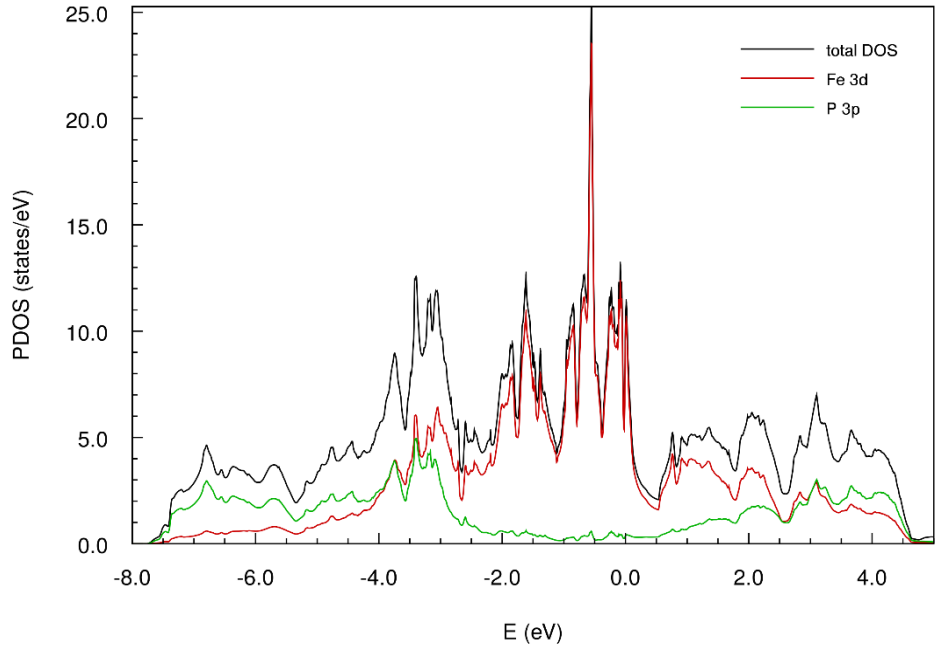

Fig. 8. Total and partial density of states for each atomic species of FeP. The Fermi level is set to 0 eV.

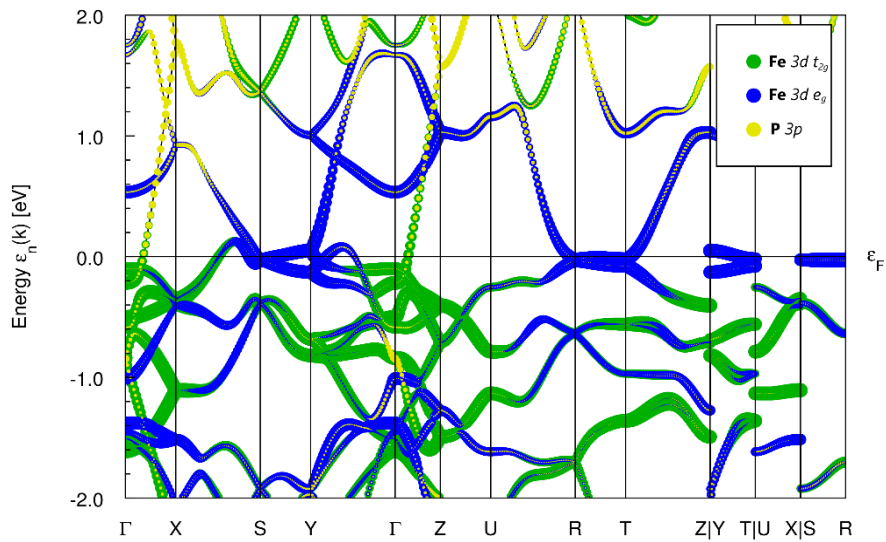

Fig. 9. Fat bands representation of the band structure of FeP. The size of the circles is proportional to the weight of 3d states of Fe and 3p states of P in the corresponding Bloch wave function.